\documentstyle[aps,prl,epsf]{revtex} 
\begin{document}
\def\SNG{{\em Physical Review Style and Notation Guide}}
\def\LUG {{\em \LaTeX{} User's Guide \& Reference Manual}}
\def\btt#1{{\tt$\backslash$\string#1}}%
\def\REVTeX{REV\TeX}
\def\AmS{{\protect\the\textfont2
        A\kern-.1667em\lower.5ex\hbox{M}\kern-.125emS}}
\def\AmSLaTeX{\AmS-\LaTeX}
\def\BibTeX{\rm B{\sc ib}\TeX}
\twocolumn[\hsize\textwidth\columnwidth\hsize\csname@twocolumnfalse%
\endcsname
\title{Long-range order versus random-singlet phases in antiferromagnetic\\
       systems with quenched disorder} 
\author{T.R.Kirkpatrick}
\address{Institute for Physical Science and Technology, and Department of Physics\\
University of Maryland, College Park, MD 20742}
\author{D.Belitz}
\address{Department of Physics and Materials Science Institute,
University of Oregon,
Eugene, OR 97403}
\date{\today}
\maketitle
\begin{abstract}
The stability of antiferromagnetic long-range order against quenched
disorder is considered. A simple model of an antiferromagnet with a
spatially varying N{\'e}el temperature is shown to possess a nontrivial 
fixed point corresponding to long-range order that is stable
unless either the order parameter or the spatial dimensionality exceeds a 
critical value. The instability of this fixed point corresponds to the 
system entering a random-singlet phase. The stabilization of long-range 
order is due to quantum fluctuations, whose role in determining the phase 
diagram is discussed.
%
%
\end{abstract}
]
 
Quantum antiferromagnetism (AFM) has experienced a surge of interest in 
recent years, both in efforts to explain the magnetic properties
of doped semiconductors\cite{R}, and
in connection with high-T$_c$ superconductivity
\cite{hiTc,SachdevChubukovSokol}. In the former context in
particular, the interplay between AFM and strong quenched
disorder is an important issue. Bhatt and Lee \cite{BhattLee} have modeled
the weakly doped, insulating regime of these systems at zero temperature
($T=0$) by an ensemble of randomly distributed, AFM
coupled Heisenberg spins with a very broad distribution of coupling
constants $J$, and employed a numerical renormalization procedure
\cite{DasguptaMa}. While, with decreasing temperature, an
increasing number of spin pairs freeze into inert singlets,
they concluded that the
remaining spins give essentially a free spin contribution
to the magnetic susceptibility. The net result is a `random-singlet' (RS)
phase, with a sub-Curie power-law behavior of the magnetic susceptibility.
The quantum nature of the spins
thus prevents the classically expected long-range order (LRO) of either
AFM or spin glass type, and explains the experimentally observed absence
of LRO. 

Bhatt and Fisher\cite{BhattFisher} have applied similar ideas
to the highly doped metallic regime. These
authors argue that rare fluctuations in the random potential always provide
traps for single electrons, which then act as randomly distributed local
moments (LM) to which the method of Bhatt and Lee can be applied, but now in a
metallic environment. Their conclusion was that the LM cannot be 
quenched by either the Kondo effect or the conduction electron induced
RKKY interaction. This leads again to a RS phase with a
magnetic susceptibility that diverges as $T\rightarrow 0$, albeit slower 
than any power. 

These results raise the important question whether, and
how, AFM LRO can {\it ever} exist in a disordered system. Intuitively,
one expects quantum fluctuations to weaken the metallic RS phase
since they enhance the interaction of the isolated electrons with their
environment. One should therefore wonder whether quantum fluctuations
can restore LRO by suppressing the RS phase that
would otherwise pre-empt an AFM transition.

\begin{figure}
\epsfxsize=7.5cm
\epsfysize=6.5cm
\epsffile{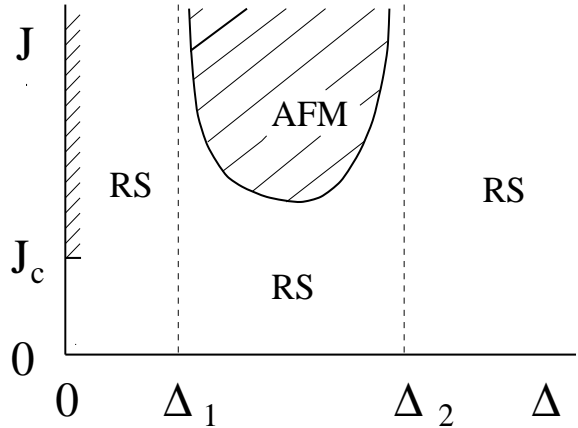}
\vskip 0.5cm
\caption{Schematic phase diagram for $d<4$ and $p<p_c$,
 showing the antiferromagnetic
 (AFM) and random-singlet (RS) phases in the $J - \Delta$
 plane. For $\Delta =0$, one has an AFM for $J>J_c$, and a Fermi
 liquid for $J<J_c$. The dependence of the phase boundary on the fluctuation
 parameter $u$ is not shown. See the text for further explanations.}
\label{fig:1}
\end{figure}

In this Letter we address these questions by studying a model of an itinerant
AFM with a spatially random N{\'e}el temperature. We first describe our main
results. We find that quantum fluctuations do indeed restore a LRO AFM
phase, provided that the order parameter dimensionality, $p$, is smaller
than a critical value $p_c$ which depends on the spatial dimensionality, $d$.
For $d=3$, we estimate $p_c >3$. The phase diagram 
in the plane spanned by the disorder, $\Delta$, and the (mean) 
AFM coupling constant, $J$, for $d<4$ and $p<p_c$ is schematically depicted in 
Fig.\ \ref{fig:1}.
In agreement with the conclusion of
Ref.\ \onlinecite{BhattFisher}, one is in the RS phase for
sufficiently small values of $J$ for all $\Delta > 0$, 
and also for sufficently small $\Delta>0$ for all values of $J$.
However, for $\Delta_1 < \Delta < \Delta_2$
there is AFM LRO for sufficiently large $J$. The transition to this
AFM state is non-Gaussian in nature, i.e. it has non-mean field exponents. 
Still larger disorder destroys
this AFM phase\cite{MIT}. At $\Delta =0$, there is the Gaussian transition
described by Hertz \cite{Hertz} from a Fermi liquid at $J<J_c$ to an AFM
at $J>J_c$. This AFM phase is unstable against
arbitrarily small amounts of disorder.

For $4<d<d^*$, where $d^*$ depends on $p$, and $p<p_c$, the phase diagram
still looks like the one shown in Fig.\ \ref{fig:1}, except that the
Gaussian AFM phase extends from the $J$-axis to finite values of
$\Delta$, up to a value $\Delta^* < \Delta_1$.
For $d>d^*$ the disorder window
$\Delta_1 < \Delta < \Delta_2$ disappears, and the transition is always
Gaussian. The non-Gaussian transition also disappears for $p>p_c$. If
$d>4$, one then has only the Gaussian transition at small $\Delta$, while
for $d<4$ there is no LRO phase.

The qualitative features of this phase diagram are due 
to the nature of the mechanism that restores the non-Gaussian transition, 
viz. quantum fluctuations that can lead to long-range spin correlations
which quench the LM. In $d>4$, disorder is much less destructive for 
LRO than in $d<4$. For small disorder, this allows for a Gaussian fixed point 
where both the quantum fluctuations and the disorder flow to zero. 
For larger disorder, quantum fluctuations are necessary to restore LRO even 
in $d>4$, and the fixed point is non-Gaussian. In $d<4$, the latter 
mechanism is the only one, so there is only a non-Gaussian transition.
The qualitative dependence of the phase diagram on the number of spin
components follows along the same lines: With
increasing $p$, as with increasing $d$, the AFM phase
shrinks. For $p>p_c$ in $d<4$, LRO can no longer be 
restored and one has a RS phase for all values of $J$. In $d>4$ for $p>p_c$
there still is a Gaussian transition, while the non-Gaussian one disappears
if either $p$ or $d$ become too large.

We have performed a perturbative renormalization group (RG)
calculation that corroborates the above ideas and results, and also sheds
additional light on the structure of the phase diagram. We also have
obtained quantitative results in the framework of a double 
$\epsilon$-expansion, working in $4-\epsilon$ space and $\epsilon_{\tau}$
imaginary time dimensions\cite{BoyanovskyCardy}. To one-loop order we
obtain for the correlation length exponent $\nu$ and the dynamical critical
exponent $z$ at the non-Gaussian transition,
\begin{mathletters}
\label{eqs:1}
\begin{equation}
\nu = {1\over 2} + {1\over 32(p-1)}\left[3p\epsilon + 4(p+2)\epsilon_{\tau}
                                    \right]\quad,
\label{eq:1a}
\end{equation}
\begin{equation}
z = 2 + {1\over 8(p-1)}\left[(4-p)\,\epsilon 
                         + 4(p+2)\,\epsilon_{\tau}\right] \quad.
\label{eq:1b}
\end{equation}
\end{mathletters}%
To this order, the exponent $\eta = 0$. All other static exponents can 
be obtained from the usual scaling laws\cite{Ma}.
In what follows we describe these explicit calculations, and show how 
they lead to the above conclusions.

Our starting point is
Hertz's action \cite{Hertz} for an itinerant quantum antiferromagnet, which is
a $\phi^4$-theory for a $p$-component order parameter field $\vec\phi$ whose
expectation value is proportional to the staggered magnetization. The bare
two-point vertex function reads,
\begin{equation}
\Gamma({\bf q},\omega_n) = t + {\bf q}^2 + \omega_n \quad.
\label{eq:2}
\end{equation}
Here $t$ denotes the distance from the critical point,
${\bf q}$ is the wavevector, and $\omega_n$ is a Matsubara frequency.
We modify this action by adding
disorder in the form of a `random mass'
term, i.e. we consider $t$ a random function of position
with a Gaussian distribution with mean $t$ and
variance $\Delta$. We use the replica trick \cite{ReplicaTrick} to 
integrate out this quenched disorder and obtain an action,
\begin{eqnarray}
S[\vec\phi^{\alpha}] = {1\over 2}
                       \int d{\bf x}\,d{\bf y}\,\int_0^{1/T} d\tau\,d\tau'\,
                       \sum_{\alpha}\ 
                       \vec\phi^{\alpha}({\bf x},\tau)\ 
\nonumber\\
                       \times\Gamma({\bf x}-{\bf y},
                       \tau - \tau')\ \vec\phi^{\alpha}({\bf y},\tau')
\nonumber\\
 + u\int d{\bf x}\,\int_0^{1/T} d\tau\,\sum_{\alpha}\left(
  \vec\phi^{\alpha}({\bf x},\tau)\cdot \vec\phi^{\alpha}({\bf x},\tau)\right)^2
\nonumber\\
 -\Delta\int d{\bf x}\,\int_0^{1/T} d\tau\,d\tau'\,\sum_{\alpha,\beta}\ \left(
  \vec\phi^{\alpha}({\bf x},\tau)\cdot \vec\phi^{\alpha}({\bf x},\tau)\right)
\nonumber\\
  \times
  \left(\vec\phi^{\beta}({\bf x},\tau')\cdot
         \vec\phi^{\beta}({\bf x},\tau')\right)\quad.
\label{eq:3}
\end{eqnarray}
Here $\alpha$ and $\beta$ are replica indices, $\tau$ denotes imaginary
time, $\Gamma({\bf x},\tau)$ is the Fourier transform of the vertex
function given in Eq.\ (\ref{eq:2}), and $u$ is the coupling constant
of the usual $\phi^4$-term \cite{Hertz}.
Note that $u,\Delta \ge 0$, so the presence of disorder
(i.e. $\Delta \neq 0$) has a destabilizing effect on the field theory.

Let us first reconsider the clean case, $\Delta =0$. We define the scale
dimension of a length $L$ to be $[L] = -1$, and that of time to be 
$[\tau] = -z$, with
$z$ the dynamical critical exponent, and look for a Gaussian fixed point
where $z=2$ and $\eta=0$. Power counting in $d$ dimensions shows
that the scale dimension of $u$ is $[u] = 4 - (d+z)$, so $u$ is irrelevant
for all $d>2$, and the Gaussian FP is stable\cite{Hertz}. In contrast,
the term $\sim\Delta$ carries an extra time integral, so with
respect to the Gaussian FP we have $[\Delta] = 4-d$. Hence the disorder
is relevant for $d<4$, and the Gaussian FP is no longer stable in the
presence of disorder. This instability of the Gaussian FP can also be
inferred from the Harris criterion\cite{Harris}.

In order to see whether there is any other FP that might be stable instead
of the Gaussian one, we have performed a one-loop RG
calculation for the model, Eq.\ (\ref{eq:3}).
A simple momentum-frequency
shell calculation yields the following flow equations,
\begin{mathletters}
\label{eqs:4}
\begin{equation}
{du\over dl} = (\epsilon - 2\epsilon_{\tau})\,u 
                              - {p+8\over 6}\,u^2 + 24\,u\,\Delta \quad,
\label{eq:4a}
\end{equation}

\begin{equation}
{d\Delta\over dl} = \epsilon\,\Delta + 16\,\Delta^2 - 
                              {p+2\over 3}\,u\,\Delta \quad,
\label{eq:4b}
\end{equation}

\begin{eqnarray}
{dt\over dl} = 2\,t - {p+2\over 6}\,u\,t + 4\Delta\,t \quad.
\label{eq:4c}
\end{eqnarray}
\end{mathletters}%
Here $l = \ln b$ with $b$ the RG length scale factor, and
we have scaled $u\rightarrow u/24$, and $\Delta\rightarrow\Delta/2$.
Our model is formally very similar to the classical model studied in
Refs.\ \cite{BoyanovskyCardy}, and our flow equations, Eqs.\ (\ref{eqs:4}),
can easily be mapped onto theirs. A
controlled loop expansion requires a double expansion in two small parameters,
which in the present case take the form of $\epsilon = d-4$, and the
number of time dimensions $\epsilon_{\tau}$. The physical case is, of course,
$\epsilon=\epsilon_{\tau}=1$, and the expansion in $\epsilon_{\tau}$ is
probably ill-behaved\cite{BoyanovskyCardy}. It is therefore reassuring to see
that the FP structure of the flow equations does not change if one formally
sets $\epsilon_{\tau}=1$, see the discussion below.

Equations (\ref{eqs:4}) allow for four FP, which we denote by 
$(u^*,\Delta^*,t^*)$. To determine the stability of the critical surface,
it suffices to discuss Eqs.\ (\ref{eq:4a},{\ref{eq:4b}). The four
FP are: (1) A Gaussian FP with $u^* = \Delta^* = 0$, 
(2) a Random FP with $u^* = 3(\epsilon + 4\epsilon_{\tau})/2(p-1)$,
$\Delta^* = [(4-p)\epsilon + 4(p+2)\epsilon_{\tau}]/32(p-1)$,
(3) an Unphysical FP with $u^* = 0$, $\Delta^* = -\epsilon/16$, and
(4) a Classical FP with $u^* = 6(\epsilon - 2\epsilon_{\tau})/(p+8)$, 
$\Delta^* = 0$.
The two FP that correspond to critical points in the present 
context are the Random and the Gaussian one. A linear stability analysis
within our one-loop calculation shows that the Gaussian FP
is stable for $d>4$, and unstable for $d<4$, independent of $p$. 
The Random FP is stable provided that $3p\epsilon > 4(p-4)\epsilon_{\tau}$. 
It thus is always stable for $d<4$ and $p<4$, which includes the physical
case $d=3$, $p=3$. For $\epsilon = \epsilon_{\tau}$, large values of $p$
will destroy the stability of the Random FP in $d<4$, while for sufficiently
small $p$ it is stable not only for $d<4$, but also in a range of $d>4$.
For $d>4$, the Random and the Gaussian FP thus may both be stable.
The eigenvalues for the Random
FP are complex, so the corrections to scaling at this
transition are oscillatory in 
nature\cite{Khmelnitskii}. The remaining two FP do not describe
phase transitions. For $d<4$, the Unphysical FP is 
inaccessable from physical, i.e. positive, bare values of $\Delta$. For
$d>4$ it is unstable. The Classical FP is the usual
stable (for $d<4$) FP for a clean, classical ($\epsilon_{\tau} =0$) system.
For a clean quantum system, it becomes unstable since the dynamical exponent
$z$ lowers the upper critical dimension, and the Gaussian FP is stable
instead. The Gaussian FP then in turn is unstable against disorder. 
By linearizing Eq.\ (\ref{eq:4c}) about the Random FP, we obtain the
correlation length exponent, Eq.\ (\ref{eq:1a}). Eq.\ (\ref{eq:1b}) and
$\eta =0$ follow from the renormalization of the terms $\omega_n$ and
${\bf q}^2$, respectively, in Eq.\ (\ref{eq:2}): The latter is not
renormalized to one-loop order, and the renormalization of the former 
yields $z=2+4\Delta^*$.

In order to determine the regions of attraction for the FP, we have 
solved the flow equations numerically for various values of $\epsilon$ 
and $\epsilon_{\tau}$. The results did not qualitatively depend on the
precise values as long as $2 \epsilon_{\tau} > \epsilon$. We restrict 
our discussion to the physical
quadrant $u>0$, $\Delta>0$. For $d<4$ and $p>p_c$ neither FP
is stable, and one finds runaway flow for all initial conditions with
$\Delta\neq 0$. This runaway flow we interpret as indicative of
LM formation, with the scale $b_c \sim \Delta^{-1/\epsilon}$ at
which $\Delta$ diverges the (mean) extension of a LM. We have obtained
further support for this interpretation by constructing a classical
instanton solution for the unstable field theory \cite{ZinnJustin}.  
The physical meaning of the instanton is a LM, and $b_c$ sets the 
scale in the instanton equation. Application of the
methods of Refs. \onlinecite{BhattLee} and \onlinecite{BhattFisher} to these
LM will then yield a RS phase. 
\begin{figure}
\epsfxsize=3.in
\epsfysize=3.in
\epsffile{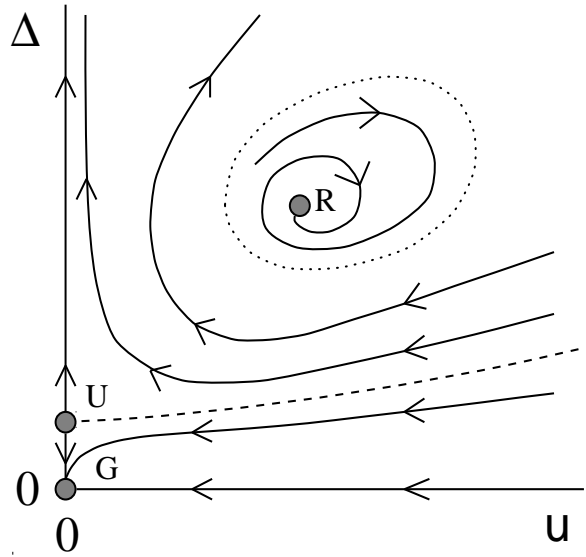}
\vskip 0.5cm
\caption{Schematic flow diagram on the critical surface for $p<p_c$ and
 $d>4$ (but $d<d^*$, see the text). The
 dotted line separates the region of attraction
 for the Random FP (R) from a region of runaway flow.
 Both the Random FP and the Gaussian FP (G)
 are stable, while the Unphysical FP (U) is unstable. The broken line is
 a separatrix that marks the limit of the region of attraction of the
 Gaussian FP.}
\label{fig:2}
\end{figure}
For $p<p_c$ (with $p_c = 16\epsilon_{\tau}/(4\epsilon_{\tau}-3\epsilon)$
in one-loop approximation), the Random FP is stable and has a finite region
of attraction. With increasing distance of the initial values from the FP
we still found the flow to be attracted by the FP, but only after oscillatory
excursions to large values of $\Delta$ and $u$, where the one-loop
approximation is no longer valid. It is natural to interpret these large
excursion as runaway flow indicative of the RS phase. This interpretation
gets support from the two-loop flow equations derived by
Boyanovsky and Cardy\cite{BoyanovskyCardy}. Solving these equations for $d<4$,
with a reinterpretation of the parameters as appropriate for the present
problem, we have found that the region of attraction for the Random FP
is finite, with a limit cycle separating the region of attraction from
a region of runaway flow. We conclude that the flow diagram on
the critical surface ($t=t^*$) is qualitatively as shown in 
Figs.\ \ref{fig:2}, \ref{fig:3}. 
For $d>4$, both the Gaussian FP and the Random FP
are stable. The former is attractive for small values of $\Delta$,
and its region of attraction is limited by a separatrix that ends in the
unstable Unphysical FP, see Fig. \ref{fig:2}. For disorder values
above that separatrix the flow is qualitatively the same as in $d<4$.
These flow diagrams correspond to the phase diagrams discussed above
and shown for $d<4$ in Fig.\ \ref{fig:1}.
We finally discuss the general structure of the phase diagram, in
particular why LRO occurs only in a disorder window.
As we have seen, the nontrivial structure of the phase diagram arises 
from competition between fluctuation induced
LRO, and LM. Technically, the instanton or LM contribution to the
free energy goes like $-\Delta^{\alpha}$, with $\alpha$ some positive
exponent. For small bare values of $\Delta$, the instanton scale
$b_c \sim \Delta^{-1/\epsilon}$ is large, and it takes many RG iterations
to reach it. The renormalized $\Delta$ is therefore
large, and the free energy gain due to the formation of LM is larger
than that due to forming LRO. The same is true for large bare values
of $\Delta$. However, for intermediate disorder values LRO is
energetically favorable. The approach to the AFM FP then leads to a
large correlation length, i.e. long-ranged spin-spin correlations
that fall off only as $1/r^{d-2+\eta}$. Such long-ranged correlations
quench the LM \cite{LRCorrelations}, which is why
the rare regions discussed in Ref.\ \onlinecite{BhattFisher} cannot
preclude the AFM transition.
\begin{figure}
\epsfxsize=3.in
\epsfysize=2.76in
\epsffile{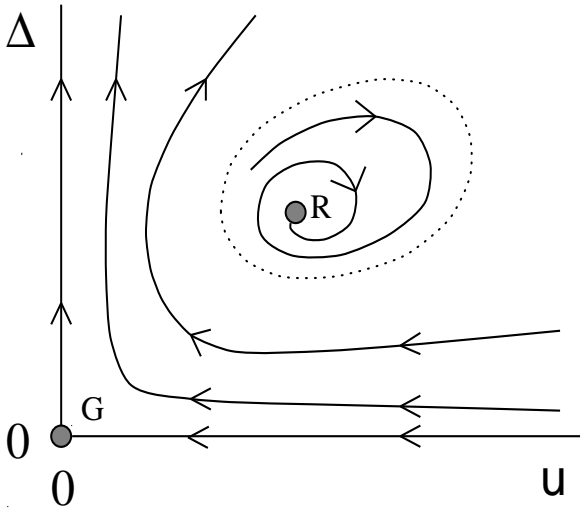}
\vskip 0.5cm
\caption{Same as Fig.\ \protect\ref{fig:2}, but for $d<4$.}
\label{fig:3}
\end{figure}

In summary, we have found that long-range
order in $3-d$ disordered AFM systems can be restored
by quantum fluctuations which destroy the
random-singlet phase. Our most striking prediction is the possibility of
re-entry into an AFM state with {\it increasing} disorder. Any attempts
to check this prediction experimentally, however, should keep in mind that
changing the disorder usually also changes other parameters, e.g. $J$ and
$u$. One might also try to confirm our
result by looking for a transition between an antiferromagnetic and
a random-singlet phase with exponents that are neither mean-field
like (as at $T=0$ in the clean case), nor classical (as at $T>0$ in
the disordered case)\cite{FiniteT}. Although the present theory has 
been formulated for itinerant electrons, this could be done 
in either metallic or insulating systems. 
 
This work was supported by the NSF under grant Nos. DMR-92-17496 and
DMR-95-10185. We greatfully acknowledge the hospitality of the TSRC in
Telluride, CO.


\begin{references}
\bibitem{FiniteT} In an actual experiment at $T>0$ one will always observe
 a crossover to classical exponents sufficiently close to the transition.
\end{references}
\end{document}